\documentclass[fleqn,twoside]{article}
\usepackage{espcrc2}
\pdfoutput=1

\usepackage{graphicx}

\newcommand{\AmS}{{\protect\the\textfont2
  A\kern-.1667em\lower.5ex\hbox{M}\kern-.125emS}}

\title{Muons in IceCube}

\author{P. Berghaus\address[UW]{University of Wisconsin, Madison (IceCube Project); Madison, WI 53703; USA} for the IceCube Collaboration\thanks{www.icecube.wisc.edu/collaboration/authorlists/2008/4.html}}
       
\begin{document}

\begin{abstract}
The IceCube detector allows for the first time a measurement of atmospheric muon and neutrino energy spectra from tens of GeV up to the PeV range. The lepton flux in the highest energy region depends on both the primary cosmic ray composition around the ``knee'' and the contribution from prompt decays of mostly charmed hadrons produced in air showers. It is demonstrated here that a direct measurement of the atmospheric muon spectrum in the region above 100 TeV is feasible using data that is already available.
\vspace{1pc}
\end{abstract}

\maketitle

\begin{figure}
\begin{center}
\includegraphics[width=2in]{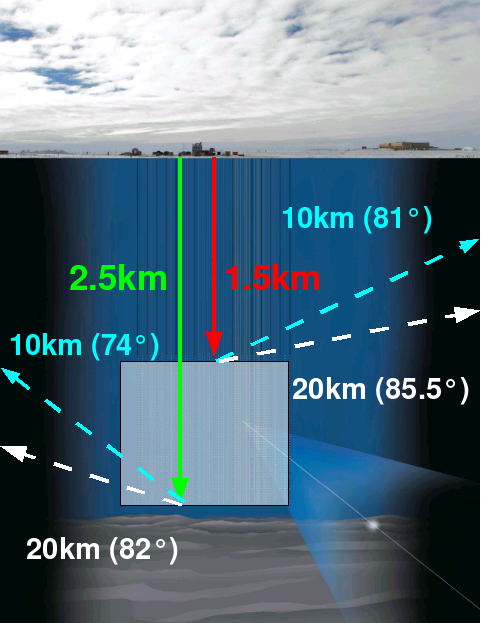}
\caption{Schematic view of the IceCube detector. The InIce neutrino detector is located from 1.5 to 2.5 km under the surface of the ice. For top and bottom of the detector, angles corresponding to slant depths of 10 km and 20 km are indicated. The region in between holds the greatest potential for a direct atmospheric muon energy spectrum measurement.}
\label{fig:detpic}
\end{center}
\end{figure}

\section{INTRODUCTION}
Neutrino astronomy is a relatively young branch of astrophysics. During the past two decades, several prototype detectors were built (Baikal, AMANDA, ANTARES). This established the basic techniques of neutrino detection in the Very High Energy (VHE) range from tens of GeV to PeV and above, but did not lead to any discovery of an astrophysical signal. 

IceCube is the first of a new generation of neutrino telescopes that will have instrumented volumes of the order of one cubic kilometer. Once completed, it is expected to put detection of astrophysical point sorces within reach \cite{Halzen:2007ip}. There are three distinct components in the IceCube array. The main volume detector {\it InIce} consists of 80 strings each holding 60 Digital Optical Modules (DOMs), located in the ice at a depth of 1.5 to 2.5 km. The surface air shower array {\it IceTop} is comprised of 160 surface tanks, and six more strings with closer-spaced DOMs will be deployed as the low-neutrino-energy extension {\it DeepCore}. A schematic drawing of the detector is shown in figure \ref{fig:detpic}. As of mid-January 2009, almost sixty strings and a similar number of surface stations have been deployed. The entire system is expected to be complete after the austral summer 2010/11 at the latest.

The main channel for neutrino detection is the process $\nu_{\mu} + N \to \mu + X$. The muon in the final state emits Cherenkov radiation when passing through ice, which can be measured using photomultiplier tubes. The direction of the muon track can then be reconstructed by means of a fit to the arrival time of the Cherenkov photons. Also, integration over the measured charge yields calorimetric information that can be used to estimate the muon energy and by extension that of the parent neutrino.

\begin{figure}
\includegraphics[width=3in]{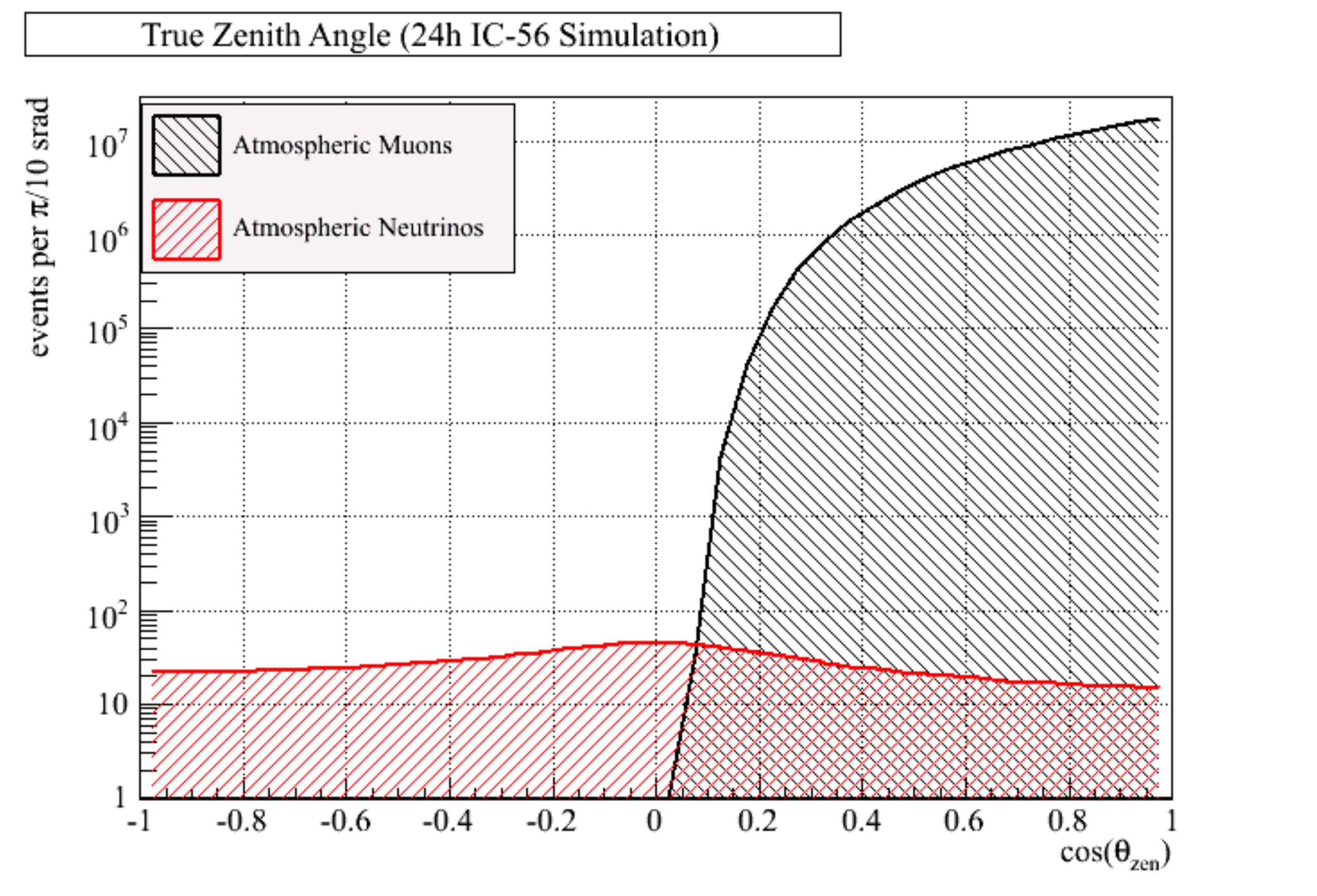}
\caption{True angular distribution of muon tracks from cosmic-ray induced air showers and atmospheric neutrinos for trigger-level Monte Carlo.}
\label{fig:muangdist_1}
\end{figure}

By far the dominant background for neutrino-induced muons comes from cosmic-ray air showers. Figure \ref{fig:muangdist_1} shows the respective event rates from atmospheric muons and atmospheric neutrinos. Above the horizon, muons dominate by several orders of magnitude. Extraction of a neutrino signal, other than at the very highest energies (tens of PeV and above), is therefore effectively only possible below the horizon, where the Earth itself can be used as a filter to attenuate cosmic-ray induced muons. 

\begin{figure}
\includegraphics[width=3in]{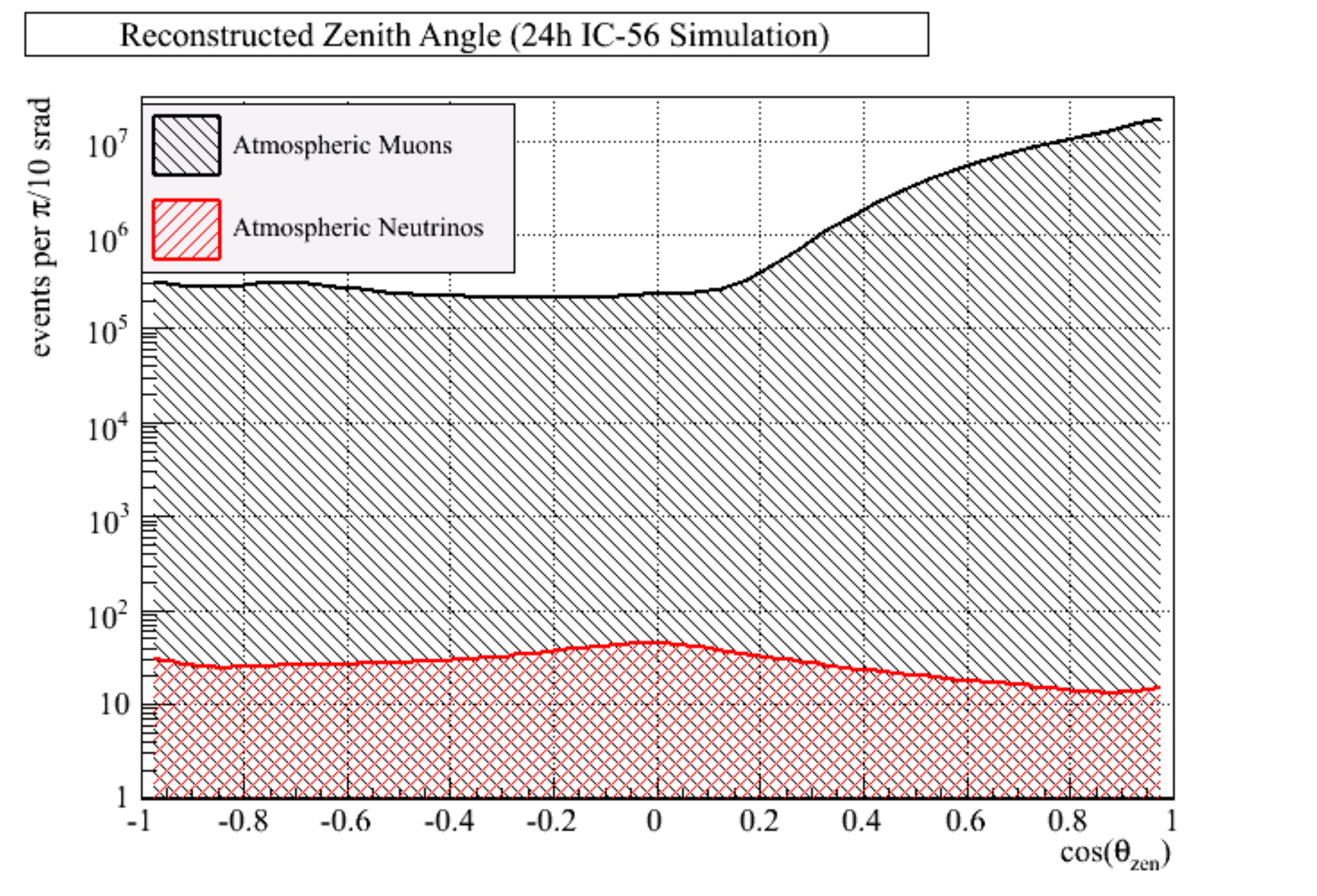}
\caption{Reconstructed muon track direction in IceCube at trigger level, before any quality cuts.}
\label{fig:muangdist_2}
\end{figure}

Even below the horizon, the background from misreconstructed downgoing muons is several orders of magnitude higher than the neutrino event rate. Figure \ref{fig:muangdist_2} illustrates the experimental situation. The main challenge for any VHE neutrino detector is to find a set of quality cuts that reduces the amount of misreconstructed muon tracks by four or five orders of magnitude while preserving as much as possible of the neutrino signal.

\section{MUON DATA IN ICECUBE}

However, the large number of atmospheric muon events registered by the detector should not simply be seen as an undesirable background for neutrino searches. In fact, it represents a high-statistics data set that can be used for calibration and verification of the detector system as well as physics investigations. 

IceCube data are transmitted to the northern hemisphere in near-real time in order to allow immediate high-level processing. Since all communication with the South Pole takes place over Low Earth Orbit satellites, the maximum data rate is restricted to about 32.5 GB/day, corresponding to roughly 6\% of the total event rate of 1400 Hz in the 40-string configuration. Data reduction is accomplished using filter conditions that make use of simple quality parameters from online event reconstructions. Each condition leads to a distinct filter stream. For atmospheric muons, the relevant data sets are:

\begin{list}{$\bullet$}{
        \setlength{\leftmargin}{25pt} \setlength{\rightmargin}{-10pt}
        \setlength{\parsep}{0pt} \setlength{\itemsep}{0pc}
        \setlength{\topsep}{2pt} \setlength{\partopsep}{0pc}
       \setlength{\parskip}{0pc}}

\item
	{\bf Muon Filter}: This stream contains all events reconstructed near or below the horizon. Its main purpose is the analysis of up-going neutrino-induced muon tracks. Event rate: tens of Hz
\item
	{\bf Minimum Bias}: For general quality control and calibration, an unbiased sample of events is provided. In the 22-string configuration (IC-22), every 100th event was stored. This has since been reduced to every 2000th. Event rate: $<$ 1 Hz
\item
	{\bf nanoDST}: For every trigger, a rudimentary set of low-level event parameters is kept. Examples are the total number of photoelectrons registered and track parameters from a basic-level reconstruction. Event rate: $>$ 1 kHz

\item
	{\bf IceTop Coincident}: This stream contains InIce events that are coincident with air showers triggering the IceTop surface array. Due to the placement of IceTop, this data sample is restricted to the area around the zenith. Event rate: tens of Hz 
\end{list}

\begin{figure}
\includegraphics[width=3in]{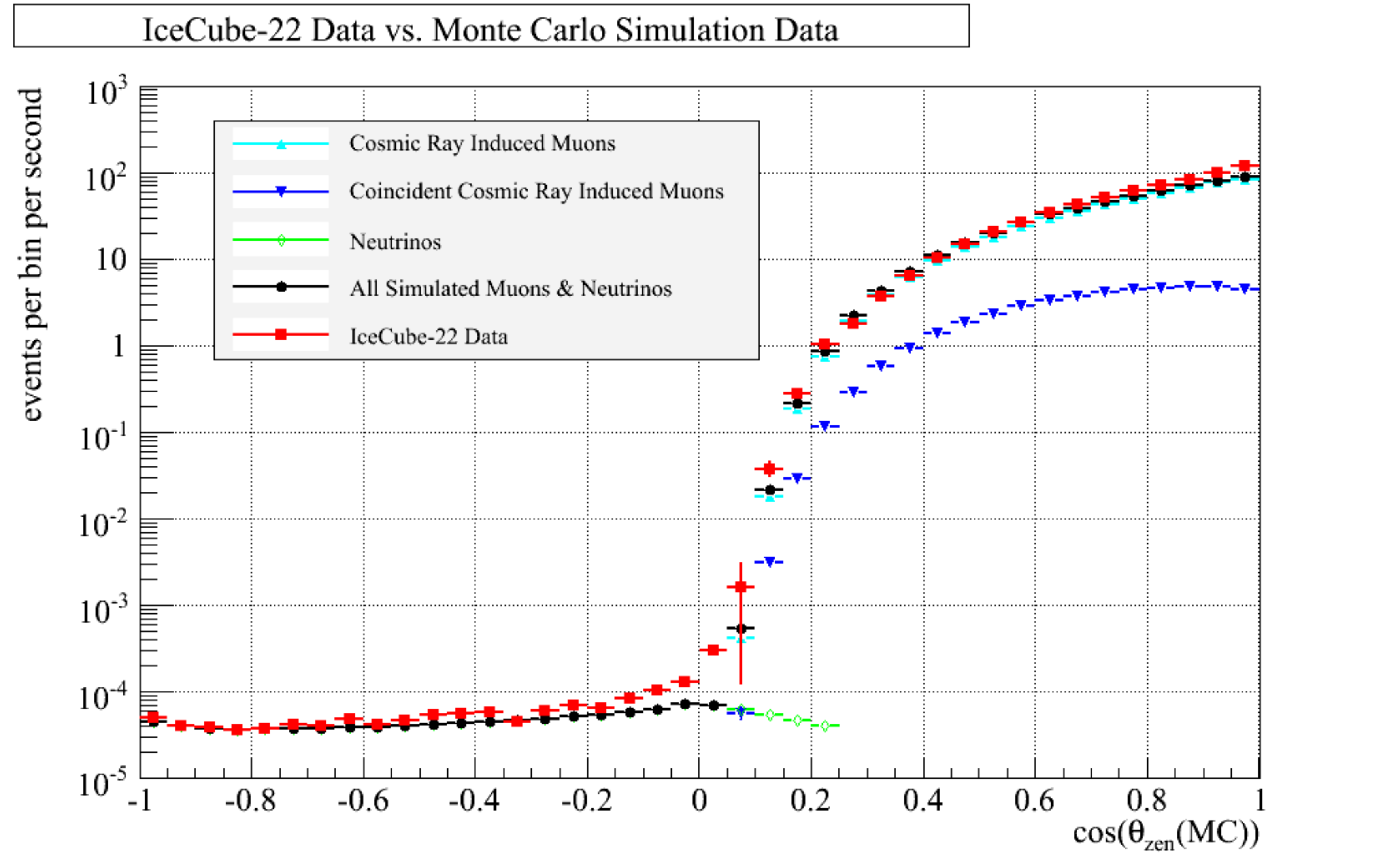}
\caption{All-Sky Muon Flux as measured with IC-22, renormalized to trigger
level (actual event rates after cuts are 8\%-20\% of the value shown).  The data in this plot represent a total livetime of 275.6 days. The downgoing muon flux has been simulated using CORSIKA/SIBYLL and is shown separately for single and multiple-primary (coincident) cosmic ray showers. Muon neutrino events were simulated up to cos $\theta_{zen}=0.25$.}
\label{fraeuleinrosenauplot}
\end{figure}

Using a combination of minimum bias and muon-filtered data, the performance of IceCube has been verified by comparing the measured muon track zenith angle distribution to a Monte Carlo simulation. Figure \ref{fraeuleinrosenauplot} shows the result, representing the first time in IceCube/AMANDA that the muon flux was measured over the entire sky using a uniform set of quality parameters. Data and simulation generally agree to better than 20\%, meaning that detector systematics are known reasonably well. A similar measurement had earlier been performed for AMANDA \cite{paolo_03}, but was restricted to angles above the horizon.

The high statistics available in the nanoDST format allows to search for possible anisotropies in the primary flux. Both the Tibet Air Shower Array and MILAGRO report a dipole anisotropy at primary energies in the TeV range \cite{Amenomori:2006bx,Abdo:2008aw}. IceCube should be able to confirm this effect based on data taken in its 22-string configuration, which took data from May 2006 to March 2007. An investigation into this issue is ongoing.

The main goal of IceTop is to measure the composition of cosmic rays around the ``knee'', located near $E_{prim} = 3\: \rm PeV$. An important parameter here is the ratio of electromagnetic to muonic particles in an air shower. This ratio can be estimated by combining information from the surface array and the InIce detector. A preliminary measurement of the primary cosmic ray spectrum has recently been completed, confirming that IceTop performs according to expectations \cite{klepser_icetop}.

\section{MUON SPECTRUM MEASUREMENT}

\subsection{Motivation}

The origin of the ``knee'' in the cosmic spectrum can be explained if the spectrum of the incident cosmic ray primaries depends on the energy per nucleon, with a sharp steepening at $E_{CR}/A \simeq 4.5\: \rm PeV$ \cite{Hoerandel:2002yg}. This change in compostion is consistent with measurements by the KASCADE array \cite{Antoni:2005wq}. Since the ratio of median parent cosmic ray and muon energy is $\leq 10$ at VHE energies \cite{gaissenthomas}, a steepening of the energy spectrum by nucleon of cosmic rays at a few PeV will have a measurable effect on the atmospheric muon spectrum at energies of hundreds of TeV.

It has also been pointed out that the muon flux measured with neutrino telescopes will be sensitive to the prompt flux from charm decay in air showers  \cite{Gelmini:2002sw}. The ratio of prompt neutrinos to prompt muons is close to unity, whereas conventional muons outnumber neutrinos by a factor of 3-5 at TeV energies, making detection of the prompt component more difficult for muons than for neutrinos. A review of prompt lepton flux calculations is given in \cite{Enberg:2008te}. AMANDA already set a limit on prompt neutrinos \cite{diffuse}. Since the prompt contribution to the atmospheric flux represents an important background in searches for a diffuse astrophysical neutrino flux, an independent limit from muons would be of particular importance to neutrino telescopes.

In order to measure any deviations from expectation, it is essential to accurately predict the spectrum of muons from light meson decays. While muon neutrinos at TeV energies mostly come from the process $K \to \nu_{\mu} + X$, for kinematical reasons muons originate predominantly in pion decays $\pi \to \nu_{\mu} + \mu$ \cite{Gaisser:2002mi}. An estimate of the pion production cross section from accelerator experiments gives an uncertainty of $15\%+12.2\% \cdot log_{10}(E_{\pi}/500\: \rm GeV)$ at $x_{lab}>0.1$ above 500 GeV \cite{Barr:2006it}. This value should also apply in good approximation to the conventional (non-prompt) muon flux. The relatively small uncertainty for the conventional flux can be used to calibrate any detector response to high-energy muons.

The energy spectrum of atmospheric muons has never been measured beyond a few tens of TeV. All results so far are consistent with, or slightly above, theoretical expectation \cite{Kochanov:2008pt}. As will be shown, IceCube should be able to extend the energy range for muon spectrum measurements by at least an order of magnitude.

\subsection{Methodology}

To be registered by InIce, muons have to traverse at least 1.5 kilometers of ice and still be energetic enough to trigger the detector. This means that there is an effective energy threshold in IceCube requiring a minimum muon energy of about 400 GeV for vertical air showers. At larger zenith angles and increasing depth, this threshold energy increases correspondingly \cite{Chirkin:2004hz}. 

\begin{figure}
\includegraphics[width=3in]{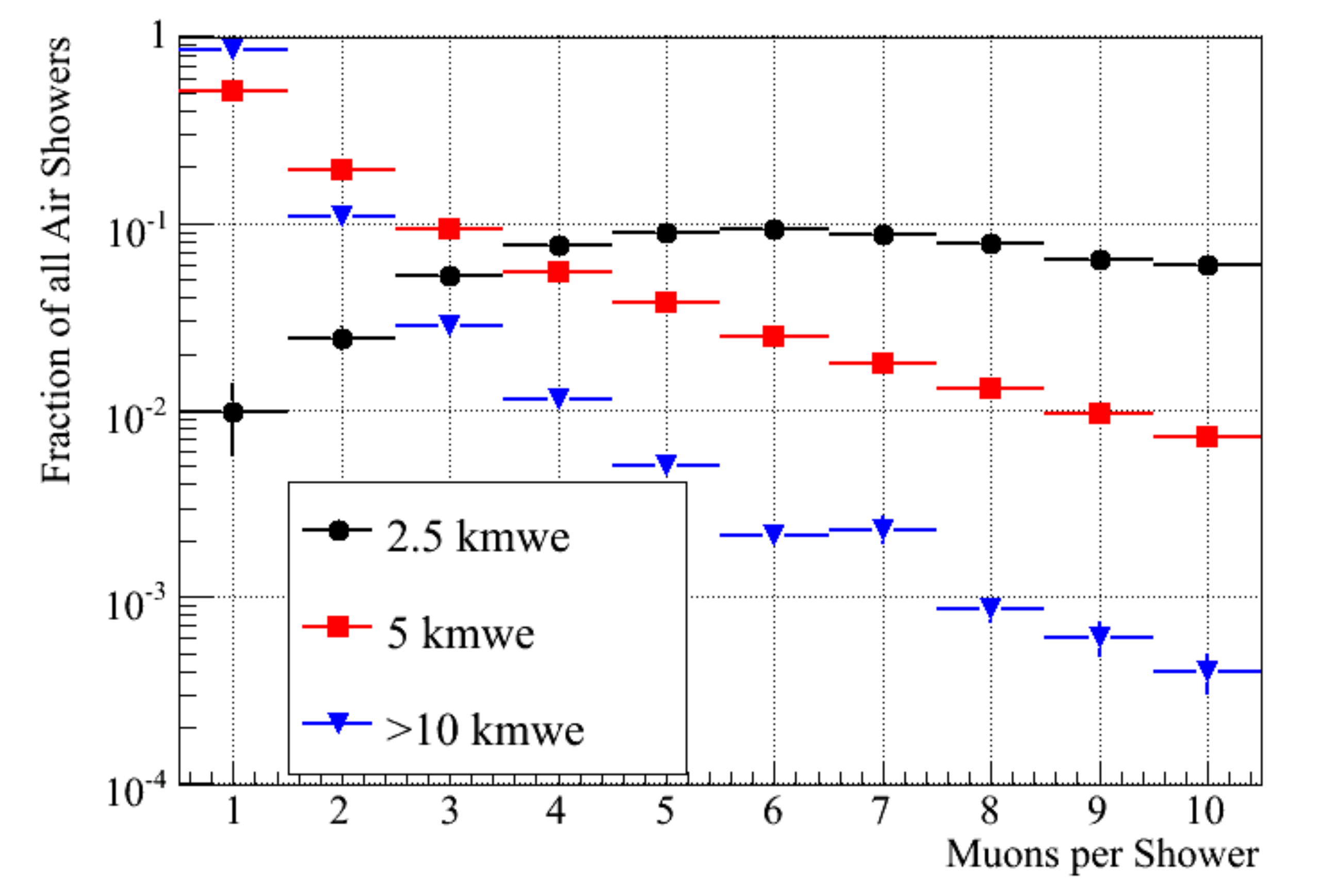}
\caption{Muon multiplicity averaged for all cosmic ray air showers at closest approach to the center of the InIce detector for three different slant depths. All figures were derived from a CORSIKA/SIBYLL-based Monte-Carlo simulation and apply to IC-22. The primary cosmic ray spectrum was parameterized according to the poly-gonato model \cite{Hoerandel:2002yg}. Propagation of muons through ice was simulated using MMC \cite{Chirkin:2004hz}.}
\label{fig:mcpix_1}
\end{figure}

Since the granularity of high-energy neutrino telescopes prevents resolution of individual muons, only the total energy of a shower can be measured. This fact has been used to derive the spectrum of the cosmic ray flux using AMANDA muon data \cite{dima_2003}. However, as shown in figure \ref{fig:mcpix_1}, the muon multiplicity decreases substantially with increasing slant depth. At a depth of 10 km water equivalent (kmwe), 90\% of all triggering air showers only contain a single muon at closest approach to the center of the InIce detector.

\begin{figure}
\includegraphics[width=3in]{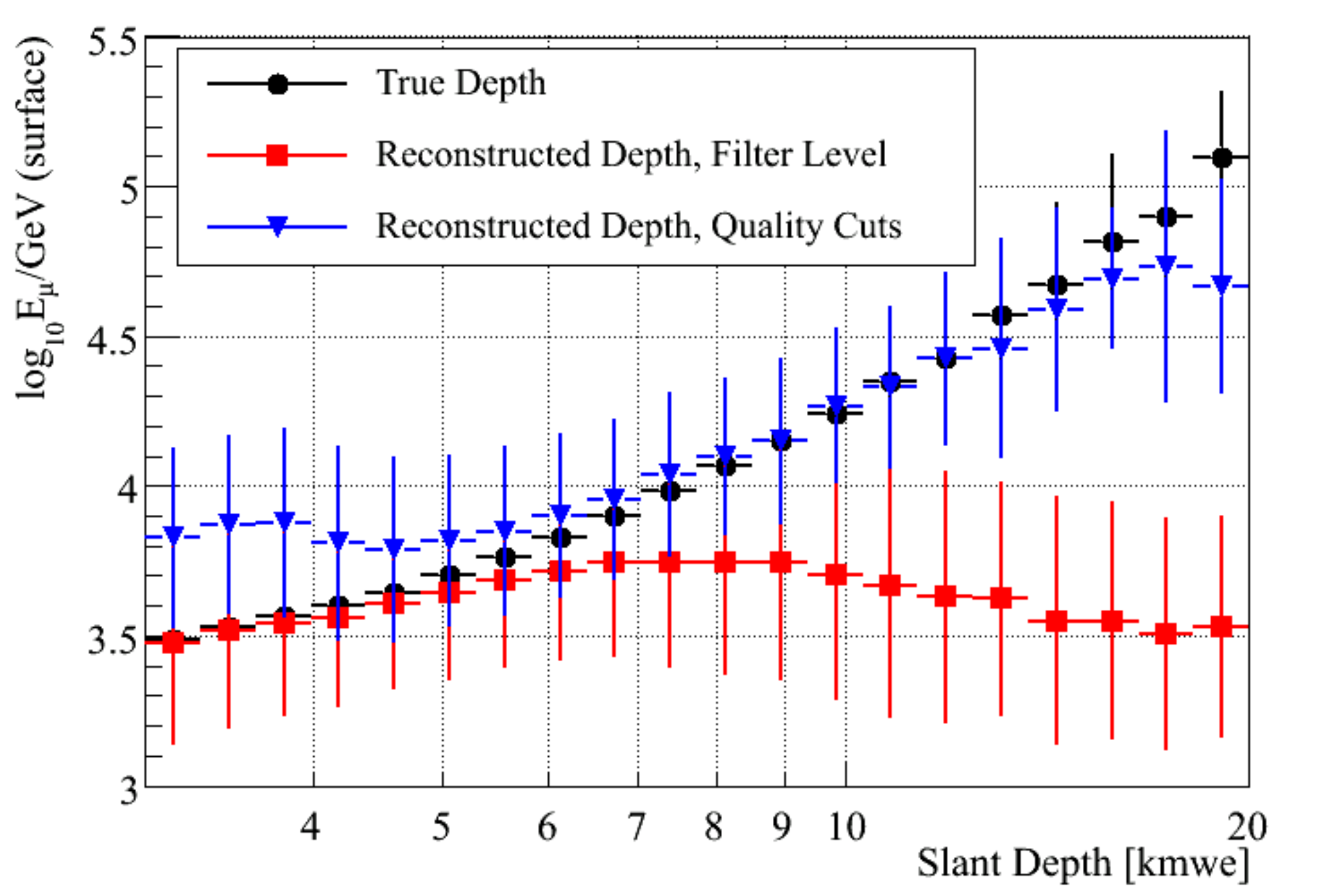}
\caption{Average surface energy of the most energetic muon in air showers depending on the slant depth measured from the surface to the point of closest approach to the center of the InIce detector.}
\label{fig:mcpix_2}
\end{figure}

\begin{figure}
\includegraphics[width=3in]{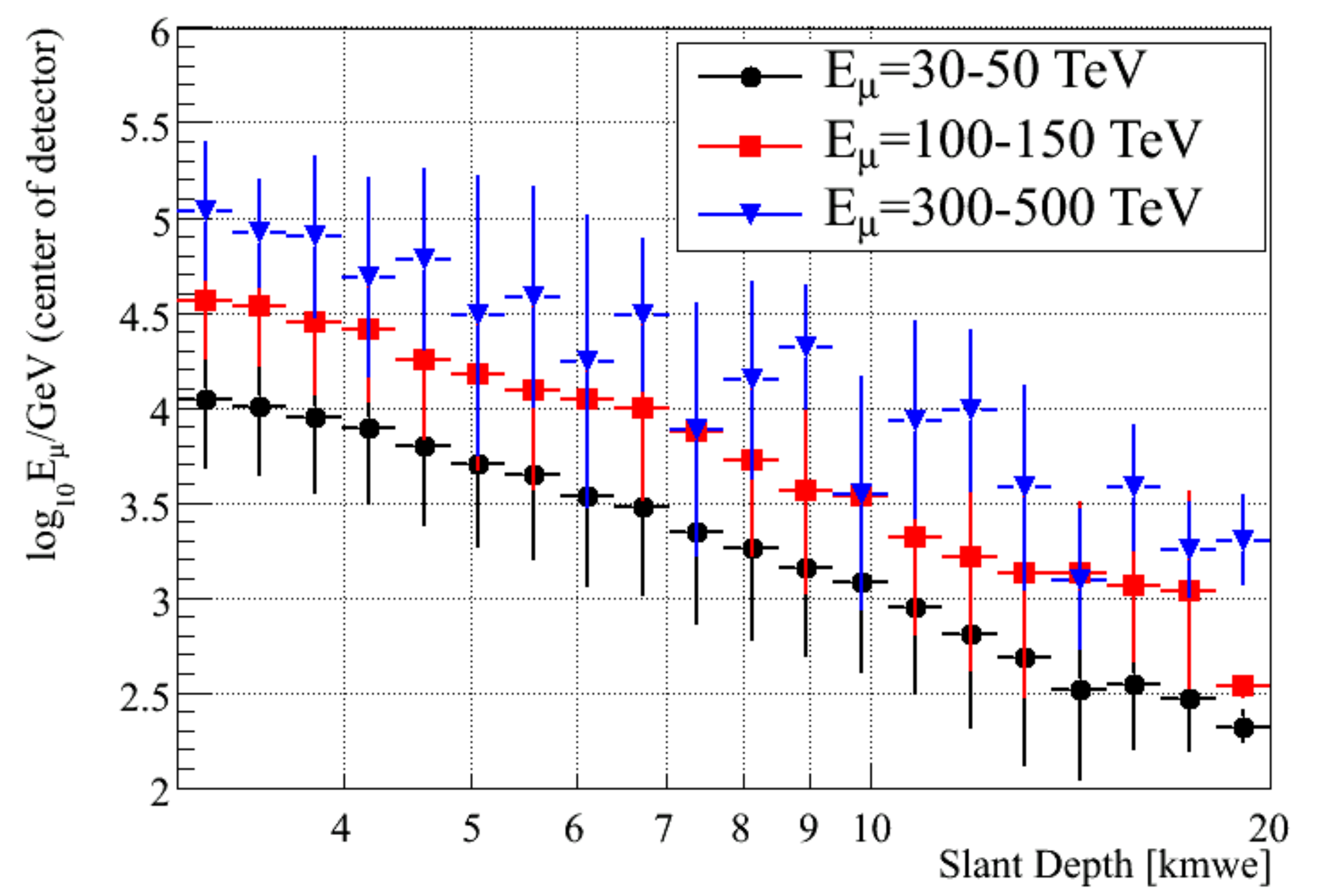}
\caption{Energy of most energetic muon in air showers at the point of closest approach to the center of the InIce detector.  The energy resolution of InIce has been estimated as 0.3 (0.5) in $log_{10}E$ for $E_{\mu}$ = 10 TeV (1 TeV) \cite{juande_dima}.}
\label{fig:mcpix_3}
\end{figure}

After a sufficiently strong quality cut, the dependence of the average muon surface energy on the measured slant depth agrees well with the true value, as shown in fig. \ref{fig:mcpix_2}. The deep edge of the region useful for a muon spectrum measurement is determined by two limiting factors. First, since the slant depth at a given zenith angle $\theta_{zen}$ is proportional to $1/cos\theta_{zen}$, finite angular resolution ($\delta \theta \approx 0.7^\circ$ in IC-22) in combination with steeply decreasing flux rates will always result in a point where the muon event rate becomes dominated by tracks from higher angles. This is illustrated by the turnover in the curve representing the measured depth after quality cuts. Second, atmospheric neutrinos become an increasingly important background, eventually drowning out the muon signal. Taken together, these two factors lead to an upper bound for the useful slant depth region of slightly less than 20 kmwe.

For an accurate energy measurement it is also necessary for the muons to retain a significant fraction of their surface energy. Muon energies can effectively only be determined for energies above 1 TeV, where the energy loss $dE_\mu/dx$ strongly depends on $E_\mu$ \cite{juande_dima}. Figure \ref{fig:mcpix_3} shows that even after passing through $>$10 km of ice, muons with $E_{surf}>100\: \rm TeV$ still fulfill this condition. Given sufficient statistics, a measurement of slant depth and energy therefore provides sufficient information to derive the muon energy spectrum.

\begin{figure} 
\includegraphics[width=3in]{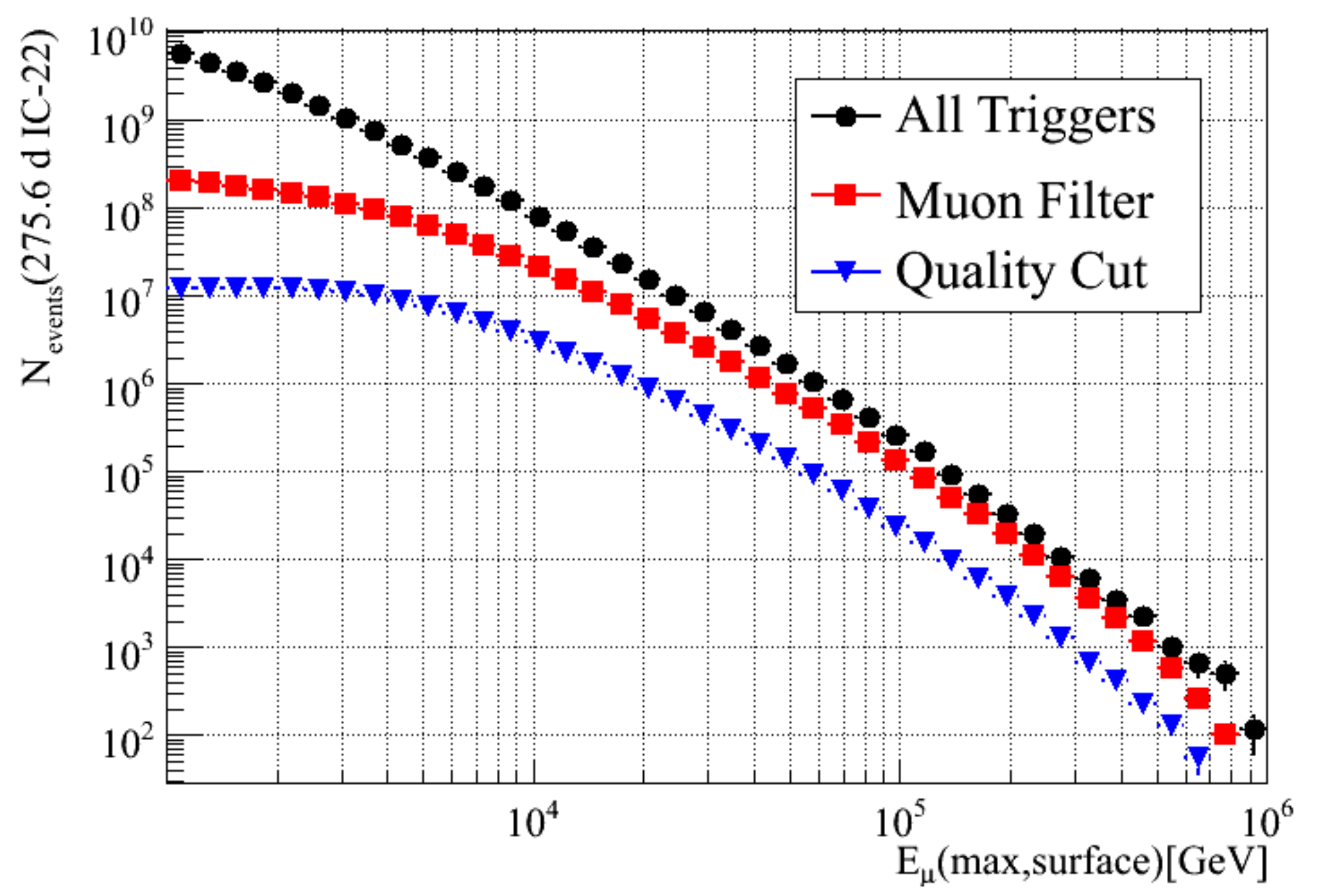}
\caption{Total IC-22 event yield from cosmic-ray induced atmospheric showers above given surface energy of the most energetic muon. The three different curves represent all IceCube triggers, events included in the muon filter event stream and after application of stringent quality cuts.}
\label{fig:evrate}
\end{figure}

Figure \ref{fig:evrate} shows the integrated event yield above a given muon surface energy for three different cut levels in IC-22. Even at this early stage in the construction of the detector, sufficient statistics should be available up to muon energies of hundreds of TeV. Since the measurement takes place at angles near the horizon, the conventional muon flux is maximized. This favors an investigation into the cosmic ray composition over a search for prompt muons.

It was further proposed to use topological separation of muon tracks belonging to the same air shower to investigate high-$p_{t} (\ge 3\: \rm GeV/c)$ muon tracks in IceCube \cite{spencer_dima}. This data can be used for an independent measurement of the cosmic ray composition which should mutually validate the result from the analysis oulined above.

\section{SUMMARY AND OUTLOOK}

Far from simply constituting a nuisance in the detection of neutrinos, atmospheric muons provide a unique opportunity to extend the scientific value of large-volume astrophysical neutrino telescopes. The combination of $\rm km^{3}$-scale detector volume and large amount of shielding material provides an opportunity to directly measure the atmospheric muon spectrum at low zenith angles in the range from tens of TeV up to 1 PeV and beyond.

Detection of a steepening of the muon spectrum in this energy region would provide independent evidence for a change in cosmic ray composition that does not rely on knowledge of the height of the shower maximum in the atmosphere, as in air shower arrays. Also, a limit on prompt muon fluxes, and hence on charm production in nuclear collisions at $E_{CM} \geq 1\: \rm TeV$ can be set. At these energies, accelerator measurements are only possible in collisions, and thus restricted to high $p_{t}$ values \cite{Acosta:2003ax}. Neutrino detectors allow investigation of the forward region and provide a natural complement to laboratory experiments.

Upcoming fixed-target experiments at the LHC will investigate light meson production at TeV energies \cite{othertalks}. In combination with results from second-generation neutrino telescopes, this should have a considerable impact on a variety of astrophysical investigations. These range from diffuse astrophysical neutrino searches to the measurement of cosmic ray fluxes at UHE energies with large air shower arrays. Further opportunities for physics investigations will undoubtedly arise as neutrino detectors increase in size and the field becomes more mature.

The analysis of IceCube data with the goal of deriving an atmospheric muon spectrum is ongoing and should yield results in the near future.

\end{document}